\RequirePackage{fix-cm}
\documentclass[smallextended,final]{svjour3}       % onecolumn (second format)
\smartqed  % flush right qed marks, e.g. at end of proof
\usepackage{graphicx}
\usepackage[a4paper, total={6.5in, 9in}]{geometry}

\usepackage{amsmath}
\usepackage{amssymb}
\usepackage{natbib}
\newcommand{\R}{\mathbb{R}}

\newcommand{\bfx}{\mathbf{x}}

\newcommand{\bfs}{{\mathbf{s}}}
\usepackage{mathrsfs}
\usepackage{xcolor}

\newcommand{\bx}{{\mathbf{x}}}
\newcommand{\bxM}{\overline{\mathbf{x}}}
\newcommand{\by}{{\mathbf{y}}}
\newcommand{\byM}{\overline{\mathbf{y}}}
\newcommand{\bz}{{\mathbf{z}}}
\newcommand{\bmu}{{\boldsymbol{\mu}}}
\newcommand{\Sig}{\Sigma^{\frac{1}{2}}}
\newcommand{\bc}{{\mathbf{c}}}

\newcommand{\tr}{{\mathrm{tr}}}

\journalname{Journal of Big Data}

\begin{document}

\title{A clustering algorithm for multivariate data streams with correlated components\thanks{This work has been partially supported by the Universit\'a degli Studi di Milano grant project 2017 "Stochastic modelling, statistics and study of the invariance properties of stochastic processes with geometrical and space-time structure in applications", and by ADAMSS Center funds for Big Data research.}}
%\subtitle{Do you have a subtitle?\\ If so, write it here}

%\titlerunning{Short form of title}        % if too long for running head

\author{Giacomo Aletti         \and
        Alessandra Micheletti %etc.
}

\authorrunning{G.~Aletti, A.~Micheletti} % if too long for running head

\institute{G. Aletti \at
              Department of Environmental Science and Policy and ADAMSS Center, \\
              Universit\`a degli Studi di Milano, Milano, Italy.\\
              Member of ``Gruppo Nazionale per il Calcolo Scientifico (GNCS)'' of the Italian ``Istituto Nazionale di Alta Matematica (INdAM)''.
              \email{giacomo.aletti@unimi.it}           %  \\
%             \emph{Present address:} of F. Author  %  if needed
           \and
           A. Micheletti \at
              Department of Environmental Science and Policy and ADAMSS Center,\\
               Universit\`a degli Studi di Milano, 
			Milano, Italy.
			\email{alessandra.micheletti@unimi.it}
}

\date{Received: date / Accepted: date}
% The correct dates will be entered by the editor

\maketitle

\begin{abstract}
Common clustering algorithms require multiple scans of all the data to achieve convergence, and this is prohibitive when large databases, with data arriving in streams, must be processed. Some algorithms to extend the popular K-means method to the analysis of streaming data are present in literature since 1998 (\cite{bradley}, \cite{O'Callaghan01streaming-dataalgorithms}), based on the memorization and recursive update of a small number of summary statistics, but they either don't take into account the specific variability of the clusters, or assume that the random vectors which are processed and grouped have uncorrelated components. Unfortunately this is not the case in many practical situations. We here propose a new algorithm to process data streams, with data having correlated components and coming from clusters with different covariance matrices. Such covariance matrices are estimated via an optimal double shrinkage method, which provides positive definite estimates even in presence of a few data points, or of data having components with small variance. This is needed to invert the matrices and compute the Mahalanobis distances that we use for the data assignment to the clusters. We also estimate the total number of clusters from the data.
% \PACS{PACS code1 \and PACS code2 \and more}
% \subclass{MSC code1 \and MSC code2 \and more}
\end{abstract}

\section{Introduction}
%\label{intro}
Clustering is the (unsupervised) division of a collection of data into groups, or \emph{clusters}, such that points in the same cluster are similar, while points in different clusters are different. 
When a large volume of (not very high dimensional) data is arriving continuously, it is impossible and sometimes unnecessary to store all the data in memory, in particular if we are interested to provide real time statistical analyses. In such cases we speak about \emph{data streams}, and specific algorithms are needed to analyze progressively the data, store in memory only a small number of summary statistics, and then discard the already processed data and free the memory \citep{GarofalakisGehrkeRastogi2016}.
Data streams are for example collected and analyzed by telecommunication companies, banks, financial analysts, companies for online marketing, private or public groups managing networks of sensors to monitor climate or environment, technological companies working in IoT, etc.  
In this framework, there are many situations in which clustering plays a fundamental role, like customer segmentation in big e-commerce web sites, for personalized marketing solutions, image analysis of video frames for objects recognition, recognition of human movements from data provided by sensors placed on the body or on a smartwatch, monitoring of hacker attacks to a telecommunication system, etc.

\subsection*{Related literature}
The methods for cluster analysis present in literature can be roughly classified into two main families: \emph{ probability-based methods} (see e.g. \cite{Aggarwal:2013:DCA:2535015}), which are based on the assumption that clusters come from a mixture of distributions, from a given family. In such case the clustering problem is reduced to the parameter estimation. These algorithms are well suited to detect the presence of non-spherical or nested clusters, but are based on specific assumptions on the data distribution, the number K of clusters is fixed at the very beginning, and, more important, they require multiple scans of the dataset to estimate the parameters of the model. Thus they can not be applied to massive datasets or data streams. 

The second family of clustering algorithms is composed by  \emph{ distance-based approaches}. Given a dataset of size $n$, grouped into $K$ clusters, such methods have usually the goal to find the $K$ \emph{centers} of the clusters which minimize the mean squared distance between the data and their closest centers. These methods  usually take different names depending on the type of considered distance. 
If the Euclidean distance is used, the corresponding method is the classical and very popular \emph{$K$-means} method (see e.g. \cite{Jain:2010}), which is probably the most diffused clustering algorithm, because of its simplicity. Anyway the exact solution of the minimization problem connected with $K$-means is NP-hard, and only local search approximations are implemented. The method is sensitive to the presence of outliers and to the initial guess for the centers, but improvements both in terms of speed and accuracy of the algorithm have been implemented in $K$-means++ \citep{Arthur:K-meanspp}, which exploits a randomized seeding technique. Unfortunately both the classical $K$-means and the $K$-means++ algorithms require multiple scans of the dataset or a random selection from the entire dataset, in order to solve the minimization problem. Since data streams can not be scanned several times and we can not (randomly) access to the entire dataset all together, also these methods are not suitable for clustering data streams.

When the elements to be clustered are not points in $\mathbb{R}^d$ but more complex objects, like functions or polygons, other clustering algorithms are used, like PAM, CLARA, CLARANS,  (\cite{Kaufman1990,CLARANS}), which are based on non-euclidean distances defined on suitable spaces. These methods are looking for \emph{medoids}, instead of means, which are the "most central elements" of each cluster and are selected from the points in the dataset. Also these algorithms  can not be efficiently applied to analyse data streams, since they either require multiple scans of the sample, or the extraction of a subsample to identify the centroids or medoids, then all data are scanned according to such identification and the medoids  are not any more updated with the information coming from the whole dataset. Actually such popular methods are suited for data which are very high dimensional (e.g. functions) or for geometrical or spatial random objects, but not for datasets with an high number of (rather small dimensional) data.

The key element in smart algorithms to treat data streams is to find methods to represent the data with summary statistics which are retained in memory, while the single data are discarded. Such summary statistics must be updated when each new observation, or group (chunk) of observations, is processed, since a second scan of the data is not allowed. 
This strategy to analyse data streams is followed in  \cite{O'Callaghan01streaming-dataalgorithms}, where the STREAM algorithm is proposed as an extension of BIRCH \citep{Zhang:BIRCH}. The STREAM method solves a so called \emph{K-Median} problem, which is a generalization of K-means where the Euclidean distance is replaced by a general distance. 
The performance of the STREAM method with respect to computational costs and quality of the clustering, measured in terms of sum of squared distances (SSQ) of data points from the assigned clusters centers, is also studied, in particular in comparison with K-means, providing good theoretical and experimental results. In the STREAM algorithm  the number K of clusters is not specified in advance, and is evaluated by an iterative combination between SSQ and the number of used centers. The main defect of the STREAM algorithm is that it uses a \emph{global} metric $D$ on the space of the data points, and thus does not take into account that different clusters may have different specific variability. Further the metric $D$ is supposed completely known and is not estimated from the data.

\begin{figure}[h]
\begin{center}
\includegraphics[scale=0.25]{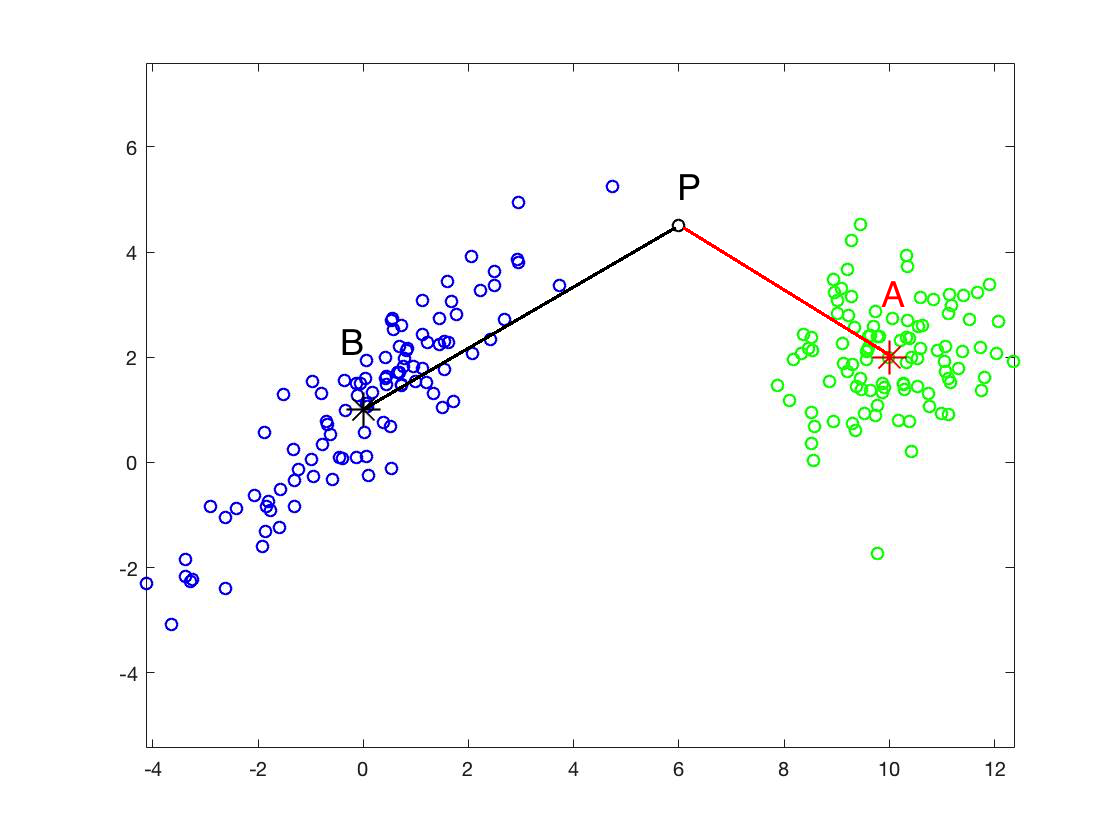}
\end{center}
\caption{A typical situation in which the shape of the point clouds must be taken into account for the points assigment: two clusters of gaussian points are represented, both composed by 100 data points. The new data point P is closer to the cluster center A if we use a global metric, like e.g. the Euclidean distance, but the point is more likely to belong to the cluster centered at B}\label{fig:ex1}
\end{figure}

In many situations the quality of the clustering is improved if a \emph{local metric} is used. A local metric is a distance which takes into account the shape of the "cloud" of data points in each cluster to assign the new points (see Figure \ref{fig:ex1}).

A first attempt to use a local distance is given by the Bradley-Fayyad-Reina (BFR) algorithm  (\cite{bradley,ullman}), which solves the K-means problem by using a distance based on the variance of each component of the random vectors belonging to the different clusters. The BFR algorithm is based on the assumption that the clusters' distribution results from a mixture of multivariate normal distributions, whose parameters are estimated from the data streams. 
The BFR Algorithm for clustering is based on the definition of three different sets of data:
\begin{itemize}
\item [a)] the \emph{retained set} (RS): the set of data points which are not recognized to belong to any cluster, and need to be retained in the buffer;
\item [b)] the \emph{discard set} (DS): the set of data points which can be discarded after updating the summary statistics;
\item[ c)] the \emph{compression set} (CS): the set of summary statistics which are representative  of each cluster.
\end{itemize}
Each data point is assigned to one of these sets on the basis of its local distance from the center of each cluster. Here the Mahalanobis distance is used, computed with respect to the sample covariance matrix of each cluster.

The main weakness of the BFR Algorithm resides in the assumption that the covariance matrix of each cluster is diagonal, which means that the components of the analyzed multivariate data should be uncorrelated.   With such assumption, at each step of the algorithm only the means and variances of each component of the clusters centers must be retained, reducing thus the computational costs. Further, in this setting the estimated covariance matrices are invertible even in presence of clusters composed just by two $p$-dimensional gaussian data points. Anyway such assumptions geometrically imply that the level surfaces (ellipsoids) of the gaussians including the data points in each cluster should be oriented with main axes parallel to the reference system. 

\subsection*{Aims and overview of the paper}

We here propose a method to clusterize data streams, using a \emph{local metric} which is estimated in real time from the data. Such metric is based on the Mahalanobis distance of the data points from each cluster center $\bc_i$, computed using an estimator of the covariance matrix of the corresponding $i-th$ cluster.
In the following we will always represent vectors as column vectors and we will assume that our data are vectors in $\mathbb{R}^p$.
\par\noindent
{\bf Definition.} Let $\bx$ be a data point and $\bc_i$ be the center of the $i-th$ cluster. Assume that the elements of the $i-th$ cluster come from a population having covariance matrix $\Sigma_i$. Then the Mahalanobis distance of $\bx$ from $\bc_i$ is given by
$$
\Delta(\bx,\bc_i)=(\bx-\bc_i)^T\Sigma_i^{-1}(\bx-\bc_i).
$$

We assume that the data points are vectors in $\mathbb{R}^d$ with correlated components and we thus estimate all the terms of the covariance matrix of each cluster, including the off diagonal ones. We use the sample mean of each cluster as centers $\bc_i$.

We divide the data in the same three sets defined in the BFR algorithm, 
we don't fix a priori the number K of clusters, and we evaluate and update such number using a density condition. Thus in our procedure from time to time new clusters will be formed, composed only by a few data points, not sufficient to obtain a positive definite estimate of the corresponding covariance matrix using the classical sample covariance estimator. We thus use an optimal double shrinkage estimator of the covariance matrix, which provides always positive definite matrices, that are then inverted to compute the Mahalanobis distance.

In our setting we will relax a little bit the assumption of gaussianity stated in the BFR algorithm, assuming that the data come from a mixture of "bell shaped" distributions, but possibly having a bigger multivariate kurtosis (i.e. fatter queues) than a gaussian.

Our algorithm is thus an improvement of the BFR algorithm, relaxing some of its assumptions.
Since with our method also the covariance terms of the clusters must be retained, there is an increase in the computational costs with respect to BFR, but such increase can be easily controlled and is affordable if the processed data are not extremely high dimensional. Therefore our algorithm is targeted to problems with data streams composed by data points of "medium" dimension, i.e. a dimension not so small to apply visualization techniques to identify the clusters (2D or 3D problems), which usually work better, but much smaller than the number of available data.

\ 

The paper is then structured as follows:  in Section \ref{shrinkage} we face the problem of the estimate of the covariance matrix of each cluster. We modify a Steinian linear shrinkage estimator in order to obtain a positive definite estimator of the covariance matrix, which can be applied also to non-gaussian cases, and which can be incrementally updated during the data processing. In Section \ref{model-update}  we introduce the summary statistics that will be retained in memory for each cluster, and we show that they can easily be updated when new data streams are processed. We then describe the way by which the data points are assigned to the three sets RS, CS, DS. In Section \ref{secondary-compression} we describe the \emph{secondary compression}, that is the way by which the points in RS and CS can be merged to pre-existing clusters or are put together to form new clusters.
In Section \ref{exp-results} we apply our method first to synthetic data, and we compare heuristically its performances with the case in which the data points are assumed to have uncorrelated components, like in the BFR algorithm. We then apply our method to cluster the real dataset KDD-CUP'99 ( \texttt{http://kdd.ics.uci.edu/databases/kddcup99/kddcup99.html}),  a network intrusion detection dataset, that was used also to test the STREAM algorithm. We apply our algorithm to all the variables in the dataset which are declared continuous. Actually some of such variables have a very small variance; anyway our optimal double shrinkage estimator of the covariance matrices of the clusters guarantees positive definite estimates also in this situation, stabilizing thus the local Mahalanobis distances that we use in our procedure. The results are coherent with the structure of the dataset, whose data should be divided into 5 clusters, as we obtain.

%It is well known that although the mean estimator used by K-means is obviously consistent if a model is considered for an isolated cluster, K-means doesn't estimate consistently
%the means of K Gaussian populations (see e.g. \cite{Bryant1978}). Anyway the bias is severe when the cluster sizes are strongly unbalanced. 

In this paper we don't study the asymptotic properties of our algorithm, but we limit ourselves to show heuristically that our algorithm provides better results of other methods to cluster data streams present in literature, just with a small increase in the computational costs.

%Actually the data points could be transformed into vectors with uncorrelated components by performing a principal component analysis (PCA). But in order to accomplish this task in our situation, where the different clusters have different covariance matrices, we would need to know a priori the cluster to which each new observation belongs, in order to use the correct covariance matrix to perform the PC transformation. We would thus need an algorithm of supervised learning, which is not our case.
%
%In the following we will describe an extension of the BFR algorithm to the case of clusters having "full" covariance matrix. 
%
% 
%
%%\section{An extension of the BFR clustering algorithm}
%
%We will use the same three sets of data a)-c) introduced in the BFR algorithm, but using  different summary statistics to define the discard set and the compression set.

\section{The covariance matrices of the clusters} \label{shrinkage}

Our algorithm is based on the Mahalanobis distance, it is hence crucial to estimate the covariance matrices of the clusters in an optimal way. Let us first observe 
 that when a new cluster is formed, it contains too few data points to obtain a positive definite estimate of the covariance matrix, using the sample covariance matrix, at least until $N\leq p$, where $N$ is the number of data in a cluster and $p$ the data points dimension.
 
To solve this problem in an optimal way, we exploit the optimal double shrinkage estimator given in \cite[Equation (3.6)]{Ikeda} by
\begin{equation}\label{eq:doubleShrink}
\hat{\Sigma}=(1-\hat{\lambda}_I-\hat{\lambda}_D)S+\hat{\lambda}_I\tfrac{tr(S)}{p}I_p+\hat{\lambda}_DD_S,
\end{equation}
where $S$ is the sample covariance matrix, $D_S$ is its diagonal matrix, $I_p$ is the identity matrix of order $p$, and $0\leq\hat{\lambda}_I+\hat{\lambda}_D\leq 1$  are weighting the convex combination of the three matrices. 
This estimator is optimal in terms of quadratic loss \citep{Himeno, Toulomis, Ikeda},
 and it leads to covariance matrix estimators that are non-singular, well-conditioned, expressed in closed form 
and computationally cheap regardless of $p$. Therefore, in these terms, it is the optimal choice
among the possible alternatives, where the first term $(1-\hat{\lambda}_I-\hat{\lambda}_D)$ 
should be initially settled close to $0$, and then its value is increasing to $1$ when $N\to \infty$.
We note that when $\hat{\lambda}_I=0$ and $\hat{\lambda}_D=1$ we obtain the local distance used in BFR.
In \citep{Ikeda}, $\hat{\lambda}_I,\hat{\lambda}_D$ are given as functions of the quantities
\begin{equation}\label{eq:defW}
\begin{pmatrix}
\lambda_I\\\lambda_D
\end{pmatrix} 
= 
\left(
\begin{smallmatrix}
\tr[(S-\tfrac{\tr(S)}{p}I_p)^2] & \tr[(S-\tfrac{\tr(S)}{p}I_p)(S−D_S)]
\\
\tr[(S-\tfrac{\tr(S)}{p}I_p)(S−D_S)] & \tr[(S-D_S)^2]
\end{smallmatrix}
\right)^{-1}  
\left(
\begin{smallmatrix}
\tr(S^2)-\widehat{\tr[\Sigma^2]}\\ \tr(S^2)-\tr(SD_S)-\widehat{\tr[\Sigma(\Sigma-D_\Sigma)]}
\end{smallmatrix}
\right),
\end{equation}
where $\widehat{\tr[\Sigma^2]}$ and $\widehat{\tr[\Sigma(\Sigma-D_\Sigma)]}$ are unbiased estimators of the corresponding quantities $\tr[\Sigma^2]$, $\tr[\Sigma(\Sigma-D_\Sigma)]$, $\Sigma$ is the true covariance matrix of the considered cluster, and $D_\Sigma$ its diagonal matrix.
Unfortunately (see \citet{Toulomis,Ikeda}), both these estimators are based on the scalar statistics
\[
Q^{(N)} = \frac{1}{N-1}\sum_{i=1}^N ((\bx_i-\bar{\bx}_N)^\top (\bx_i-\bar{\bx}_N))^2,
\]
proposed by \citep{Himeno}, where 
\begin{equation}\label{def:xM^N}
\bxM_N = \frac{1}{N}\sum_{n=1}^{N} \bx_n,
\end{equation}
is the centroid of the considered cluster, composed by $N$ data points.
In data stream framework, we note that $Q^{(N)}-Q^{(N-1)}$ is not a function of 
few summary statistics, which can be updated when a new data point is added to the cluster. 
In fact, $\bar{\bx}_N$ is changing with $N$ and  $Q^{(N)}$ must be then recomputed, due to the quadratic term in its definition, using all the data in the cluster when a new point is added.
To overcome this problem, we prove in the following section the existence of two unbiased estimators for
$tr[\Sigma^2]$ and $tr[\Sigma(\Sigma-D_\Sigma)]$ based on the following statistics $Q_N$:
\begin{equation}\label{def:Q_N}
Q_N = 
\begin{cases}
\big( (\bx_2  - \bx_1 )^{\top} (\bx_2  - \bx_1 ) \big)^2 & \text{if }N=2;\\
Q_{N-1} +  
\big( (\bx_N  - \bxM_{N-1})^{\top} (\bx_N  - \bxM_{N-1}) \big)^2 & \text{if a new point } \bx_N \text{ is added} \\
&  \text{to a cluster of }N-1\text{ points};\\
\ & \ \\
Q_{N_1} + Q_{N_2}  
& \text{if a cluster is made by merging} \\
& \text{two clusters of }N_1\text{ and }N_2\text{ points}.\\
\end{cases}
\end{equation}
The key point is that  $Q_N$ is defined 
recursively, and it is a function of $Q_{N-1}$, the new added point, and the centroid of the cluster at the time of the update.

%We use the diagonal matrix $D_S$ of the sample covariance matrix $S$ as  ``target
%matrix'' of the shrinkage method, noting that $D_S$ was the BFR estimate of the covariance of each cluster used in \cite{bradley}.
%In other words, in presence of few data, our method coincides with that of \cite{bradley}, and we allow a progressive influence of
%correlation as the number of data increases.

In the following we will describe the details of our method 
%to determine the optimal value of $\lambda$ 
and the assumptions that must be satisfied to apply it.

\subsection{A model for the estimate of the covariance matrices}
Our dataset is given by a sequence of $p$-dimensional vectors $\bx_1, \bx_2, \ldots$.
Each observation $\bx_n$ is independent on the others and, if belonging to the cluster ${\underline{k}}$, it is generated as
\[
\bx_n = {\bmu}_{\underline{k}} + \Sig_{\underline{k}} \bz_n
\]
where ${\bmu}_{\underline{k}}$ is the mean vector and $\Sig_{\underline{k}} $ is a matrix such that
$\Sigma_{\underline{k}} = \Sig_{\underline{k}} (\Sig_{\underline{k}})^{\top}$ is strictly positive definite.
The following hypothesis of uncorrelation is assumed on the first four moments:
\begin{equation}\label{eq:model}
E[ \bz_n] = \mathbf{0}, \qquad \mathrm{Cov}(\bz_n ) = E[ \bz_n \bz_n^{\top}] = \mathbf{I}, \qquad
E[\prod_{i=1}^q z_{n,i}^{\gamma_i}] = \prod_{i=1}^q E[z_{n,i}^{\gamma_i}] ,
\end{equation}
for any integers $\gamma_1, \ldots, \gamma_q$ satisfying $0\leq \sum_{1}^q \gamma_i \leq 4$, and where
$z_{n,i}$ is the $i$-th component of the vector $\bz_n = (z_{n,1}, \ldots, z_{n,q})^{\top}$.

Assume that the sequence $\bx_1, \bx_2, \ldots$ belongs to the same cluster with $\Sig_{\underline{k}} = \Sig$.
Then the sequence $\by_1, \by_2, \ldots$ defined as
\(\by_n = \bx_n - \bmu_{\underline{k}} = \Sig \bz_n \),
is formed by independent vectors with null expectation.
Then, as a consequence of \eqref{eq:model}, we have that 
\[
E [ \by_i^{\top} \by_j ] = 
\begin{cases}
E[ \bz_i^{\top} (\Sig)^{\top} \Sig \bz_i ] = 
\tr ( (\Sig)^{\top} \Sig) = %\tr (\Sig (\Sig)^{\top} )= 
\tr (\Sigma)
& \text{if }i = j;
\\
0 & \text{otherwise}.
\end{cases}
\]
Moreover, 
\(
E [ \by_i^{\top} \by_j \by_k^{\top} \by_l] \neq 0
\)
only in the following situation:
\begin{subequations}\label{eq:Ey4}
\begin{description}
\item[when $i=j=k=l$] then
\begin{equation}\label{eq:Eyyyy}
E [ \by_i^{\top} \by_j \by_k^{\top} \by_l] =
 E[ (\by_i^{\top}\by_i)^2] = \kappa_{11} + 2 \tr (\Sigma^2) + (\tr \Sigma)^2,
\end{equation}
where $\kappa_{11}$ is defined in \cite{Himeno} as 
$$
\kappa_{11}\colon = E[\bz_i'\Sigma \bz_i \bz_i'\Sigma\bz_i]-2\tr (\Sigma^2) - (\tr \Sigma)^2.
$$
Note that $\kappa_{11}=0$ for gaussian data, thus it is an indicator of deviation from gaussianity in terms of kurtosis. In case of gaussian data its estimation can be neglected \citep{fisher}.
\item[when $(i=j)\neq(k=l)$] then
\begin{equation}\label{eq:Eyyzz}
E [ \by_i^{\top} \by_j \by_k^{\top} \by_l] =
E[ (\by_i^{\top}\by_i) (\by_k^{\top}\by_k) ]
=E[ (\by_i^{\top}\by_i)] E[ (\by_k^{\top}\by_k) ]
=
(\tr \Sigma)^2;
\end{equation}

\item[when $(i=l)\neq(j=k)$] then
\begin{equation}\label{eq:Eyzzy}
\begin{aligned}
E [ \by_i^{\top} \by_j \by_k^{\top} \by_l] &=
E[ \bz_i^{\top} (\Sig)^{\top} \Sig ( \bz_j \bz_j^{\top} ) (\Sig)^{\top} \Sig \bz_i ] \\
&= E[ \bz_i^{\top} (\Sig)^{\top} \Sig E[ \bz_j \bz_j^{\top} ] (\Sig)^{\top} \Sig \bz_i ] \\
&=
E[ \bz_i^{\top}  (\Sig)^{\top} \Sig  (\Sig)^{\top} \Sig \bz_i] \\& 
= \tr ((\Sig)^{\top} \Sig  (\Sig)^{\top} \Sig )= 
\tr (\Sig (\Sig)^{\top} \Sig  (\Sig)^{\top} )= \tr (\Sigma^2 );
\end{aligned}
\end{equation}

\item[when $(i=k)\neq(j=l)$] the same as above, since $\by_k^{\top} \by_l= \by_l^{\top} \by_k$, hence
\begin{equation}\label{eq:Eyzyz}
E [ \by_i^{\top} \by_j \by_k^{\top} \by_l] =
\tr (\Sigma^2 ).
\end{equation}
\end{description}
\end{subequations}

\begin{lemma}\label{lem:EyM2}
As a consequence of \eqref{eq:Eyzyz},
\[
E \Big[ \Big( \by_N^{\top}\sum_{i=1}^{N-1} 
\by_i \Big)^2 \Big] = (N-1) \tr (\Sigma^2) .
\]
%%% BEGIN COLOR RED
%{\color{red}%
%\[
%E \Big[ \Big( \by_n^{\top}\sum_{i=1}^{n} 
%\by_i \Big)^2 \Big] = (n+1) \tr (\Sigma^2)+
%\kappa_{11} +  (\tr \Sigma)^2 .
%\]
%}
%%% END COLOR RED
\end{lemma}

\begin{lemma}\label{lem:EMMMM}
As a consequence of all the relations \eqref{eq:Ey4},
\[
E \Big[ \sum_{i,j,k,l=1}^{N-1} 
\by_i^{\top} \by_j \by_k^{\top} \by_l \Big] = (N-1)\kappa_{11} + 2(N-1)^2 \tr (\Sigma^2) + (N-1)^2(\tr \Sigma)^2.
\]
\end{lemma}

\begin{lemma}\label{lem:EyyMM}
As a consequence of \eqref{eq:Eyyzz},
\[
E \Big[ \sum_{i,j=1}^{N-1} \by_N^{\top}\by_N
\by_i^{\top} \by_j \Big] 
%= \sum_{i,j=1}^{N-1} 
%E \Big[ \by_N^{\top}\by_N
%\by_i^{\top} \by_j \Big] 
= (N-1) (\tr \Sigma)^2 .
\]

%%% BEGIN COLOR RED
%{\color{red}%
%\[
%E \Big[ \Big( \sum_{i,j=1}^{n} \by_n^{\top}\by_n
%\by_i^{\top} \by_j \Big)^2 \Big] = n (\tr \Sigma)^2 +
%\kappa_{11} + 2 \tr (\Sigma^2) .
%\]
%}
%%% END COLOR RED
\end{lemma}

\subsection{Optimal shrinkage estimation}

We now use the previous results to solve the problem of finding the optimal estimates of 
$\hat{\lambda}_I,\hat{\lambda}_D$ in \eqref{eq:doubleShrink}, as a function of the 
statistics $S$ (sample covariance matrix of the data in the same cluster), of $Q_N$ given in \eqref{def:Q_N}, and of
two quantities $\mathbb{S}_N$ and $\mathbb{T}_N$ that can be updated inductively.
%The optimal weight for the nonparametric linear shrinkage estimation, 
%in terms of minimizing the risk function relative to the quadratic loss,
%is computed as follows (see, e.g., \cite{fisher,Himeno,Toulomis,Ikeda})
%\[
%\frac{E[\tr( S(S - D_S ))] - E [\tr( \Sigma (S - D_S ))]}{ E[\tr ((S - D_S )^2) ]}
%\]
%It is obvious that $E[\tr( S(S - D_S ))]$ and $E[\tr ((S - D_S )^2) ]$
%may be directly unbiasedly estimated by $\tr( S(S - D_S ))$ and $\tr ((S - D_S )^2) $ respectively. 
As can be seen in \eqref{eq:defW}, the problem here is the unbiased estimation of the terms 
\(\tr[\Sigma^2]\) and \( \tr( \Sigma^2 ) - \tr ( D_\Sigma^2) .\) The derivation of this estimate is given
in the next section, after a technical result given hereafter.
%In \cite{Toulomis,Ikeda}, the estimation is based on the statistic $Q$ given for the first time in \cite{Himeno} as
%\[
%Q = \frac{\sum_{n=1}^{N}  \big( (\bx_n  - \bxM)^{\top} (\bx_n  - \bxM) \big)^2}{N-1}
%\] to obtain, e.g.\ in
%\cite[Eq. (2.17)]{Ikeda} 
%\[
%\hat{\lambda} = 0 \vee
%\frac{\tr ( S^2 )/p -\tr ( S D_S )/p - \big( \widehat{\tr( \Sigma^2 ) - \tr ( D_\Sigma^2)} \big)/p }{ \tr ( S^2 )/p - (\tr  S /p)^2 } \wedge 1.
%% \frac{E[\tr( S(S - D_S ))] - E [\tr( \Sigma (S - D_S ))]}{ E[\tr ((S - D_S )^2) ]}
%\]
%The statistics $Q$ was added to correct a bias due to non-gaussian models. 

We may use the following
additional relations in our estimates \citep{Himeno,Ikeda}%to estimate $\tr( \Sigma^2 ) - \tr ( D_\Sigma^2)$,
\begin{subequations}\label{sys:old}
\begin{align}%\label{eq:eq2}
E[\tr ( S^2 ) ] & = 
\frac{1 }{N} \kappa_{11} + \frac{N }{N-1} \tr (\Sigma^2) + \frac{1 }{N-1} (\tr \Sigma)^2 
\\
E[(\tr S)^2 ] & = 
\frac{1 }{N} \kappa_{11} + \frac{2 }{N-1} \tr (\Sigma^2) +  (\tr \Sigma)^2 
\\
E[\tr(D_S^2)] & = %E[ \tr( S D_S) ] = 
\frac{1}{N-1} \kappa_{11} + 
\frac{N+1}{N-1} \tr (D_\Sigma^2) + 
\frac{R_N}{N-1},
%\\
%E[Q_N] & = \kappa_{11}  \sum_{n=2}^N\Big(1 + \frac{1 }{(n-1)^3} \Big)+ 
%( 2\tr (\Sigma^2) + (\tr \Sigma)^2 ) \sum_{n=2}^N\Big(1 + \frac{1 }{n-1} \Big)^2
\end{align}
\end{subequations}
once we have recalled that the quantity $R_N$ is negligible (see, again, \cite{Toulomis,Ikeda}).
When the data are distributed as gaussians,
a direct estimation without $\kappa_{11}$ based on (\ref{sys:old}a-c) may be done (see \cite{fisher}), since $\kappa_{11} =0$. 
%Unfortunately, the statistic $Q$ cannot be updated by keeping in memory only
%simple summary statistics when two clusters are merged together or when a single point is
%added to a cluster. 
%With this problem in mind, we have defined the modification of $Q$, that can be simply updated, while, 
%on the other hand, we loose the
%exchangeability property of $Q$.
When this is not the case, we may use the statistics $Q_N$ already introduced in \eqref{def:Q_N} and
we will prove in Lemma~\ref{lem:EQ_N} that
\begin{equation}\label{eq:EQ_N}
E[Q_N] = \mathbb{S}_N \kappa_{11}  + 
\mathbb{T}_N ( 2\tr (\Sigma^2) + (\tr \Sigma)^2 ) ,
\end{equation}
where $\mathbb{S}_N$ and $\mathbb{T}_N$ are two quantities that may be simply calculated inductively as:
\begin{align}\label{def:S_N}
\mathbb{S}_N &=
\begin{cases}
2 & \text{if }N=2;\\
\mathbb{S}_{N-1} + \Big(1 + \frac{1 }{(N-1)^3} \Big)
& \text{if a new point is added} \\
&  \text{to a cluster of }N-1\text{ points};\\
\  & \ \\
\mathbb{S}_{N_1} + \mathbb{S}_{N_2}  
& \text{if a cluster is made by merging} \\
&  \text{two clusters of }N_1\text{ and }N_2\text{ points};\\
\end{cases}
\\ \label{def:T_N}
\mathbb{T}_N &=
\begin{cases}
4 & \text{if }N=2;\\
\mathbb{T}_{N-1} + \Big(1 + \frac{1 }{(N-1)} \Big)^2
& \text{if a new point is added} \\
& \text{to a cluster of }N-1\text{ points};\\
\  & \  \\
\mathbb{T}_{N_1} + \mathbb{T}_{N_2}  
& \text{if a cluster is made by merging} \\
& \text{two clusters of }N_1\text{ and }N_2\text{ points};\\
\end{cases}
\end{align}
%
%Define
%\[
%\mathbb{S}_N = \sum_{n=2}^N\Big(1 + \frac{1 }{(n-1)^3} \Big), 
%\qquad \mathbb{T}_N = \sum_{n=2}^N\Big(1 + \frac{1 }{n-1} \Big)^2
%\]

% and hence 

%\section{$\bxM^{(n)}$ independent of $\bx_n$}

\begin{lemma}\label{lem:EQ_N}
With the notations of \eqref{def:xM^N}, \eqref{def:Q_N},    \eqref{def:S_N} and \eqref{def:T_N} we have
\[
E[Q_N] =
\begin{cases}
2 \kappa_{11}  + 
4 ( 2\tr (\Sigma^2) + (\tr \Sigma)^2 )  & \text{if }N=2;\\
E[Q_{N-1}] +  \big(1 + \frac{1 }{(N-1)^3} \big) \kappa_{11}  
 & \text{if a new point is added} \\
\qquad + 
(1 + \frac{1 }{N-1} )^2 ( 2\tr (\Sigma^2) + (\tr \Sigma)^2 )
& \qquad \text{to a cluster of }N-1\text{ points};\\
E [Q_{N_1}] + E[Q_{N_2}]
& \text{if a cluster is made by merging} \\
& \qquad \text{two clusters of }N_1\text{ and }N_2\text{ points};\\
\end{cases}
\]
and hence
\[
E[Q_N] = \mathbb{S}_N \kappa_{11}  + 
\mathbb{T}_N ( 2\tr (\Sigma^2) + (\tr \Sigma)^2 ).
\]
\end{lemma}
See the Appendix for the proof.

\subsection{Unbiased estimators of $\tr(\Sigma^2) $ and $\tr( \Sigma^2 ) - \tr ( D_\Sigma^2)$}
Let ${\bf X} = (\tr(S^2),(\tr S)^2, \tr(D_S^2), Q_N)^\top$ and 
${\bf Y} = (\kappa_{11}, \tr (\Sigma^2) , (\tr \Sigma)^2, \tr(D_\Sigma^2))^\top$. 
We are interested in an unbiased estimator of the vector
\[
{\bf Z}=
\begin{pmatrix} 
\tr(\Sigma^2) \\ 
\tr( \Sigma^2 ) - \tr ( D_\Sigma^2)
\end{pmatrix} 
= 
B {\bf Y} ,
\text{ where } B =   
\begin{pmatrix} 
0 & 1& 0 & 0 
\\ 
0 & 1& 0 & -1
\end{pmatrix} .
\]
The system composed by (\ref{sys:old}a-c) and  \eqref{eq:EQ_N} may be read as
\[
E({\bf X}) = A {\bf Y} , \text{ where } A =   \begin{pmatrix} 
  \frac{1 }{N}  & \frac{N }{N-1} & \frac{1 }{N-1} & 0 \\ 
  \frac{1 }{N}  & \frac{2 }{N-1} & 1 & 0 \\ 
  \frac{1 }{N-1}  & 0 & 0 & \frac{N+1}{N-1} \\ 
  \mathbb{S}_N & 2 \mathbb{T}_N & \mathbb{T}_N & 0
  \end{pmatrix}.
\]
Now, the matrix $A$ may be shown to be invertible, and hence $\hat{{\bf Z}} = B A^{-1} {\bf X}$ is a linear (in ${\bf X}$) unbiased
estimator for ${\bf Z}$, since
\[
E(\hat{{\bf Z}}) = E(B A^{-1} {\bf X}) = B A^{-1} E({\bf X}) = B A^{-1} A{\bf Y} = B {\bf Y} = {\bf Z}.
\]
For sake of completeness, we give here the elements of the matrix $B A^{-1} = [C_{kl}]_{{k=1,2}\atop{l=1, \ldots,4}}$.
Let $K = (N + 2 +\tfrac{2}{N-1})\mathbb{S}_N - 3\mathbb{T}_N $, we have
\begin{align*}
C_{11} 
& = %\Big( 
\frac{  (N - 1)(N\mathbb{S}_N - \mathbb{T}_N) }{ K(N-2) }
%\Big)
;
\\
C_{12} 
& = %\Big( 
\frac{  N\mathbb{S}_N - (N-1)\mathbb{T}_N }%
{ K(N-2) }
%\Big)
;
\\
C_{13} & =
0;
\\
C_{14} & = %\Big( 
\frac{  1 }%
{ K }
%\Big)
;\\
C_{21} 
& = %\Big( 
\frac{ (N+ 1 + \frac{2}{N-2}) \mathbb{S}_N 
-( 3 +\frac{1}{N-2} - \frac{2}{N+1}) 
\mathbb{T}_N }{ K};
%\Big);
\\
C_{22} & = %\Big( 
\frac{ - ( 1 + \frac{2}{N-2} )
\mathbb{S}_N + (\frac{1}{N - 2} +\frac{1}{N+1} )\mathbb{T}_N 
}{ K}
%\Big)
;
\\
C_{23} & =
- 1+\frac{2}{N+1};
\\
C_{24} & = 
%\Big( 
\frac{\tfrac{1}{N-1} }{ K}
%\Big)
.
\end{align*}

\section{Summary statistics and primary data compression} \label{model-update}

In this section we define the summary statistics that will be retained in memory for each cluster and we describe the first phase of our clustering procedure.
As in the BFR algorithm, we first perform the \emph{primary data compression}, that is the identification of items which can be assigned to a cluster, and then discarded (Discard Set, DS), after updating the corresponding summary statistics contained in the Compression Set CS.  Data compression refers thus to representing groups of points by their summary statistics and purging these points from RAM.
In our algorithm, like in BFR, primary data compression will be followed by a secondary data-compression, which takes place over data points in the Retained Set (RS), not compressed in the primary phase.

Assume that data points $\bfx_1,\dots ,\bfx_N\in\R^p, N\geq 2$ must be compressed in the same cluster. We will retain only the following summary statistics
\begin{equation}
\Sigma_N=\sum_{i=1}^N\bfx_i \bfx_i^\top,\quad \bfs_N=\sum_{i=1}^N\bfx_i,\quad N, \label{summarystat1}
\end{equation}
and the statistics $Q_N, \mathbb{S}_N, \mathbb{T}_N$ defined in (\ref{def:Q_N}),(\ref{def:S_N}),(\ref{def:T_N}), respectively.

In particular the statistics $\bfs_N$ are needed to compute the sample means $\bar{\bfx}_N=\frac{1}{N}\sum_{i=1}^N\bfx_i$, that are used as clusters centers, while the matrices $\Sigma_N$ are used to compute the unbiased sample covariance matrices of the clusters $S=\frac{1}{N-1}\sum_{i=1}^n(\bfx -\bar{\bfx}_N)(\bfx -\bar{\bfx}_N)^\top$, which are needed, together with $Q_N, \mathbb{S}_N, \mathbb{T}_N$,  to compute the optimal double shrinkage estimators described in the previous section.

The summary statistics (\ref{summarystat1}) can also be easily updated when a new data point $\bfx_{N+1}$ must be added to the cluster, without processing again the already compressed points. In fact
$$
\Sigma_{N+1}=\Sigma_N+\bfx_{N+1} \bfx_{N+1}^\top,\quad \bfs_{N+1}=\bfs_N+\bfx_{N+1},
$$
while the other summary statistics have already been defined recursively.

Note that the matrix $\Sigma_N$ is symmetric, thus at each step of the algorithm we have to retain in memory only 
$\frac{p(p+1)}{2}+p+4=\frac{p^2}{2}+\frac{3}{2}p+4$ 
summary statistics for each cluster, where $p$ is the dimension of the data points. Thus, in case of $K$ clusters, our computational costs are of the order of $Kp^2$. In addition, note that we should simply sum the corresponding statistics if we want to merge
two clusters.

\ 

%%%%%

Similarly to the BFR algorithm, in order to assign a point to a cluster we use the squared Mahalanobis distance from its center (sample mean), i.e. we assign a new data point $\bfx$ to cluster $h$ with center $\bar{\bfx}_h$ and estimated covariance matrix $\hat{S}_h$, if $h$ is the index which minimizes
$$
\Delta^2_{\hat{S}_h}(\bfx,\bar{\bfx}_h)=(\bfx-\bar{\bfx}_h)^T (\hat{S}_h)^{-1}(\bfx-\bar{\bfx}_h).
$$
%and if $\Delta^2_{\hat{S}_h}(\bfx,\bar{\bfx}_h)$ is smaller than a fixed threshold $\delta$. 
Differently from the BFR algorithm, here we estimate the covariance matrices of the clusters with the optimal double shrinkage estimators described in the previous section. 
In order to avoid the inversion of a matrix and thus to reduce the computational costs, we observe that the Mahalanobis distance between two points $\bx, \by$, computed with respect to a covariance matrix $S$, can be rewritten as follows (see e.g. \cite[Expression A.7.10]{Rencher}):
\begin{equation}
\Delta^2_{S}(\bx,\by) = (\bx-\by)^T S^{-1} (\bx- \by) = \frac{\det[S +(\bx- \by)(\bx-\by)^T]}{\det(S)} -1 
\label{dist_quad}
\end{equation}
In our algorithm we will actually use expression (\ref{dist_quad}) for the computation of all the Mahalanobis distances.

We also compare $\bfx$ with each point $\bfx_o$ in
the retained set RS, if any, by computing
$$
\Delta^2_{\hat{S}_P}(\bfx,\bfx_o)=(\bfx-\bfx_o)^T (\hat{S}_P)^{-1}(\bfx-\bfx_o),
$$
where $\hat{S}_P$ matrix is the pooled covariance matrix based on $\hat{S}_h$ of all the $K$ clusters:
\begin{equation}\label{eq:pooled}
\hat{S}_P = \frac{ n_{h_1} \hat{S}_{h_1} + n_{h_2} \hat{S}_{h_2} + \cdots n_{h_K} \hat{S}_{h_K} }{ n_{h_1}+ n_{h_2} + \cdots + n_{h_K}},
\end{equation}
and where $n_{h} $ is the number of points in cluster $h$.
With $\hat{S}_P$, we emphasize the weighted importance of directions that are more significant for the clusters when we compute the distance
between two ``isolated'' points. Since the retained set contains the points which do not belong clearly to one specific cluster, with this comparison we check if they can be aggregated with the new incoming data, to form new clusters.
\begin{figure}[th]
\begin{center}
\includegraphics[scale=0.25]{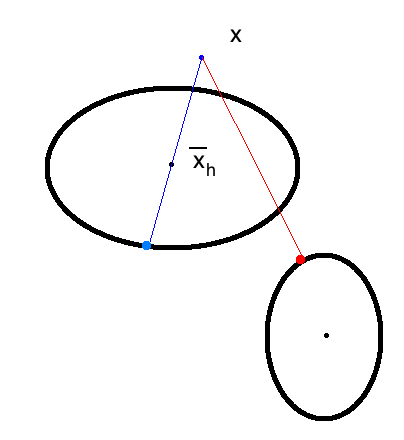}
\includegraphics[scale=0.25]{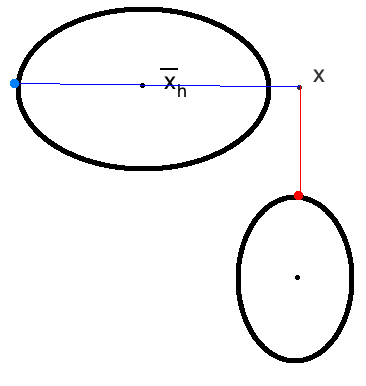}
\end{center}
\caption{The 2 possible situations after the perturbation of the clusters centers. Left: the point $\bfx$ is assigned to cluster $h$;  right: $\bfx$ is moved to RS}\label{fig:confreg}
\end{figure}

\begin{figure}[h]
\begin{center}
\includegraphics[scale=0.25]{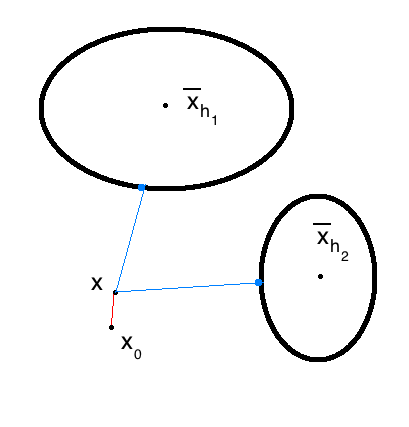}
\includegraphics[scale=0.25]{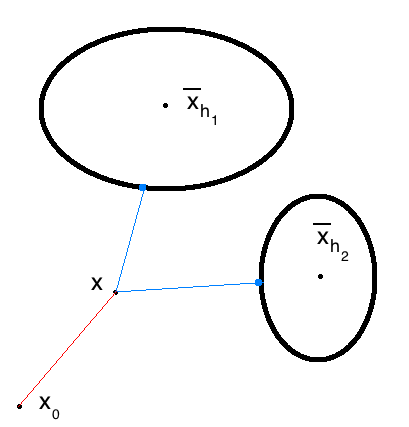}
\end{center}
\caption{The two possible situations, after the centers' perturbation, when $\bfx$ is closer to a point $\bfx_o$ of the retained set in the first comparison. Left: the points $\bfx$ and $\bfx_o$ are joined to form a new cluster;  right: $\bfx$ and $\bfx_o$ are moved to RS}\label{fig:confpoints}
\end{figure}

We then approximate locally the distribution of the clusters with a $p-$variate Gaussian and we build  confidence regions around the centers of the clusters (see \cite{Hotelling}). Following the approach stated in \cite{bradley}, which is motivated by the assumption that the mean is unlikely to move outside of the computed confidence interval, we perturb $\bar{\bfx}_h$ by moving it in the farthest position  from $\bfx$ in its confidence region, while we perturb the centers of the other clusters by moving them in the closest positions with respect to $\bfx$ and we check if the cluster center closer to $\bfx$ is still $\bar{\bfx}_h$. If yes, we assign $\bfx$ to cluster $h$, we update the corresponding summary statistics and we put $\bfx$ in the discard set; 
 otherwise, we put $\bfx$ in the retained set (RS) (see Figure \ref{fig:confreg}). 
 If in the first comparisons the point $\bfx$ is closer to a point $\bfx_o$ of the retained set than to any cluster, we form a new secondary cluster with the two points if $\bfx_o$ remains the closest to $\bfx$ after the centers' perturbation. In this case we add the corresponding summary statistics to the compressed set CS, and we put $\bfx$ and $\bfx_o$ in the discard set. Otherwise we put $\bfx$ and $\bfx_o$ in RS (see Figure \ref{fig:confpoints}).

%In other words, data points belonging to RS are candidate to be either outliers, or centers of subsequent secondary clusters.
Let us see the procedure of centers' perturbation in deeper detail.
\subsubsection{Confidence regions}

It is well-known (\cite{Hotelling}) that a confidence region for the mean $\bmu$ based on $\bxM$ and $\hat{S}$ may be based on the 
Hotelling's $T$-squared distribution
\[
t^2 = n(\bxM-\bmu)^{\top}{S}^{-1}(\bxM-\bmu)
 \sim T^2_{p,n-1}=\frac{p(n-1)}{n-p} F_{p,n-p} ,
\]
where $F_{p,n-p}$ is the F-distribution with parameters $p$ and $n-p$.

Then, if we denote by $CI_{\underline{k}}$ the confidence region for the mean of cluster ${\underline{k}}$, i.e.\
\[
CI_{\underline{k}} = \{\bmu\colon n (\bxM_{\underline{k}}-\bmu)^{\top}{\hat{S}_{\underline{k}}}^{-1}
(\bxM_{\underline{k}}-\bmu)
\leq T^2_{p,n-1}(1-\alpha) \}
\]
then the perturbation $p_{\underline{k}}(\bx)$ for the data point $\bx$ is
\[
p_{\underline{k}}(\bx) = 
\begin{cases}
\sup_{\bmu\in CI_{\underline{k}}}  (\bx-\bmu)^{\top}{\hat{S}_{\underline{k}}}^{-1}(\bx-\bmu) & \text{if }\underline{k} = j;
\\
\inf_{\bmu\in CI_{\underline{k}}}  (\bx-\bmu)^{\top}{\hat{S}_{\underline{k}}}^{-1}(\bx-\bmu) & \text{if }\underline{k} \neq j;
\end{cases}
\]
Denoting by $t_\alpha =  T^2_{p,n-1}(1-\alpha) $, if we
introduce a   Lagrange multiplier $\lambda^*$, the problems of minimization or maximization stated in the definition of $p_{\underline{k}}(\bx)$ can be solved by differentiating the following lagrangian form $\mathcal{L}$:
\[
\mathcal{L} (\bmu, \lambda) = 
(\bx-\bmu)^{\top}{\hat{S}_{\underline{k}}}^{-1}(\bx-\bmu) - n\lambda^* \big(
(\bxM_{\underline{k}}-\bmu)^{\top}{\hat{S}_{\underline{k}}}^{-1}
(\bxM_{\underline{k}}-\bmu) -\tfrac{t_\alpha}{n} \big) .
\]
The resolution $\nabla_\bmu  \mathcal{L} = \mathbf{0}$ gives 
\(
\bmu = \frac{\bx-\lambda \bxM_{\underline{k}}}{1-\lambda}
\),
where $\lambda = n \lambda^*$.
In particular, the optimal $\bmu$ is the linear combination of $\bx$ and $\bxM_{\underline{k}}$ in $CI_{\underline{k}}$ which is
farther from $\bx$ or  closer to $\bx$, when $\underline{k}=j$ or $\underline{k}\neq j$, respectively.
The constrain reads
\[
 (\bxM_{\underline{k}}-\frac{\bx-\lambda \bxM_{\underline{k}}}{1-\lambda})^{\top}{\hat{S}_{\underline{k}}}^{-1}
(\bxM_{\underline{k}}-\frac{\bx-\lambda \bxM_{\underline{k}}}{1-\lambda}) = \frac{t_\alpha}{n}
\quad \Longrightarrow \quad 
\frac{t_\alpha}{n} = 
 \frac{ (\bxM_{\underline{k}}-\bx)^{\top}{\hat{S}_{\underline{k}}}^{-1}
(\bxM_{\underline{k}}-\bx)}{(1-\lambda)^2}.
\]
Denoting by $\Delta^2_{\underline{k}, \bx} = (\bxM_{\underline{k}}-\bx)^{\top}{\hat{S}_{\underline{k}}}^{-1}
(\bxM_{\underline{k}}-\bx)$, we have 
\(
\lambda = 1 \pm \sqrt{n\Delta^2_{\underline{k}, \bx}/t_\alpha}
\) and 
\[
p_{\underline{k}}(\bx) = (\bx-\frac{\bx-\lambda \bxM_{\underline{k}}}{1-\lambda})^{\top}{\hat{S}_{\underline{k}}}^{-1}
(\bx-\frac{\bx-\lambda \bxM_{\underline{k}}}{1-\lambda}) = 
\frac{\lambda^2}{(1-\lambda)^2}  \Delta^2_{\underline{k}, \bx}.
\]
Summarizing we obtain the following perturbations of the clusters centers, referred to the data point $\bx$,
\[
p_{\underline{k}}(\bx) = 
\begin{cases}
(\sqrt{\Delta^2_{\underline{k}, \bx} } + \sqrt{ t_\alpha/n  } )^2
& \text{if }\underline{k} = j;
\\
(\sqrt{\Delta^2_{\underline{k}, \bx} } - \sqrt{ t_\alpha /n } )^2
 & \text{if }\underline{k} \neq j\text{ and }\Delta^2_{\underline{k}, \bx}\geq t_\alpha  .
\\
0
 & \text{if }\underline{k} \neq j\text{ and }\Delta^2_{\underline{k}, \bx}< t_\alpha  .
\end{cases}
\]

\section{Secondary data compression}\label{secondary-compression}

The purpose of secondary data compression is to identify ``tight'' sub-clusters of points among the data that we can not discard in the primary phase. In \cite{bradley} this is made using  the euclidean metric.
%Finally, the number of clusters is initialized to $K$, and it can increase or decrease during the procedure.

We adopt a similar idea, but we use a local metric, based on the Mahalanobis distance. We exploit
%Then, a hierarchical clustering is performed using the Ward's method \citep{GanExMurtagh,Legendre2}: 
%the distance between two clusters $h_1$ and $h_2$ 
%with $n_{h_1}\geq 2,n_{h_2}\geq 2$ points and centroids $\bar{\bfx}_{h_1}$ and $\bar{\bfx}_{h_2}$, is given by
%\[
%\Delta^2 (h_1,h_2) = \frac{n_{h_1}n_{h_2}}{n_{h_1}+n_{h_2}}  (\bar{\bfx}_{h_1} - \bar{\bfx}_{h_2})^{\top} \hat{S}_P^{-1} (\bar{\bfx}_{h_1} - \bar{\bfx}_{h_2}).
%\]
%where $\hat{S}_P=\frac{n_{h_1}\hat{S}_{h_1}+n_{h_2}\hat{S}_{h_2}}{n_{h_1}+n_{h_2}-2}$ is the pooled sample covariance matrix of the two clusters.
a technique based on hierarchical clustering, mimicking the Ward's method \citep{GanExMurtagh,Legendre2}.
%: 
%the distance between two clusters $h_1$ and $h_2$ 
%with $n_{h_1}\geq 2,n_{h_2}\geq 2$ points and centroids $\bar{\bfx}_{h_1}$ and $\bar{\bfx}_{h_2}$, is given by
%\[
%\Delta^2 (h_1,h_2) = \frac{n_{h_1}n_{h_2}}{n_{h_1}+n_{h_2}}  (\bar{\bfx}_{h_1} - \bar{\bfx}_{h_2})^{\top} \hat{S}_P^{-1} (\bar{\bfx}_{h_1} - \bar{\bfx}_{h_2}).
%\]
%where $\hat{S}_P=\frac{n_{h_1}\hat{S}_{h_1}+n_{h_2}\hat{S}_{h_2}}{n_{h_1}+n_{h_2}-2}$ is the pooled sample covariance matrix of the two clusters.
%The distance between a single retained point $\bx$ and a cluster $h$ is computed by the squared Mahalanobis distance between the point and the cluster centroid, 
%named $\Delta^2(\bx,h)$ based on the estimated covariance matrix of the cluster, while the distance between two retained points is computed by their squared Mahalanobis distance based on the pooled covariance matrix (\ref{eq:pooled}) of all the clusters.

Given two clusters $h_1$ and $h_2$ 
with $n_{h_1}\geq 2,n_{h_2}\geq 2$ points,  and centroids $\bar{\bfx}_{h_1}$ and $\bar{\bfx}_{h_2}$, respectively,
then the squared Mahalanobis distance 
of one centroid to the other cluster may be measured as $\Delta^2_{\hat{S}_{h_i}}(\bar{\bfx}_{h_1},\bar{\bfx}_{h_2})$, $i=1,2$.
Accordingly, to decide whether two clusters are close or not, we compare the weighted combination of those distances
\[
\Delta_{h_1,h_2}^2 = \frac{\tr(\hat{S}_{h_1})}{\tr(\hat{S}_{h_1})+\tr(\hat{S}_{h_2})}\Delta^2_{\hat{S}_{h_1}}(\bar{\bfx}_{h_1},\bar{\bfx}_{h_2})+
\frac{\tr(\hat{S}_{h_2})}{\tr(\hat{S}_{h_1})+\tr(\hat{S}_{h_2})}\Delta^2_{\hat{S}_{h_2}}(\bar{\bfx}_{h_1},\bar{\bfx}_{h_2}),
\]
with the the squared Mahalanobis distance 
$\Delta^2_{\hat{S}_{h_1h_2}}(\bar{\bfx}_{h_1},\bar{\bfx}_{h_2})$ of the two centroids, evaluated with the pooled  covariance matrix of the two clusters
%\[
%\Delta^2 (h_1,h_2) = \frac{n_{h_1}n_{h_2}}{n_{h_1}+n_{h_2}}  (\bar{\bfx}_{h_1} - \bar{\bfx}_{h_2})^{\top} \hat{S}_P^{-1} (\bar{\bfx}_{h_1} - \bar{\bfx}_{h_2}).
%\]
%where 
$\hat{S}_{h_1h_2}=\frac{n_{h_1}\hat{S}_{h_1}+n_{h_2}\hat{S}_{h_2}}{n_{h_1}+n_{h_2}}$.
The distance between a single retained point $\bfx$ and a cluster $h$ is computed by the squared Mahalanobis distance 
$\Delta^2_{\hat{S}_h}(\bfx,\bar{\bfx}_h)$ between the point and the cluster centroid, based on the estimated covariance matrix of the cluster, while the distance between two retained points $\bfx_1,\bfx_2$ is computed by their squared Mahalanobis distance $\Delta_{\hat{S}_{P}}^2(\bx_1,\bx_2)$ 
based on the pooled covariance matrix (\ref{eq:pooled}) of all the clusters.

Based on the hierarchical tree built with such distances, we sequentially merge two clusters or points only if a suitable density condition is fulfilled.
This condition is different for the different types of merging that we can perform:
\begin{itemize}
\item we merge two clusters $h_1$ and $h_2$ if $\Delta_{h_1,h_2}^2 <\theta_0 \Delta^2_{\hat{S}_{h_1h_2}}(\bar{\bfx}_{h_1},\bar{\bfx}_{h_2})$;
\item we merge a retained point $\bx$ and a cluster $h$ if $\Delta^2_{\hat{S}_h}(\bfx,\bar{\bfx}_h)<\theta_1(tr(S_{h}))$;
\item we merge two retained points $\bx_1$ and $\bx_2$ if $\Delta_{\hat{S}_{P}}^2(\bx_1,\bx_2)<\theta_2$.
\end{itemize}

Here $\theta_i$, $i=0,1,2$, are thresholds, chosen by the user. 
%In our case study, we used $\theta_0 = 1$, $\theta_1$ corresponding to a  proportion of  the total variances of the groups and
%$\theta_0$ a fixed value. 
For what concerns $\theta_2$, we suggest to use a significant quantile of the $\chi$-square distribution that arises under the null hypothesis \\
$H_0$: the retained points
come from a gaussian distribution with covariance matrix given by the pooled covariance matrix (\ref{eq:pooled}) of all the clusters.
 
\section{Results on simulated and real data and discussion}\label{exp-results}

\subsection{Results on synthetic data}
Synthetic data were created for the cases of 5 and 20 clusters. Data were sampled from 5 or 20 independent $p$-variate Gaussians, with elements of their mean vectors (the true means) uniformly distributed on $[-5,5]$. 
The covariance matrices were generated by computing products of the type
$\Sigma=UHU^T$, where $H$ is a diagonal matrix with elements on the diagonal distributed as a $Beta(0.5,0.5)$ rescaled to the interval  $[0.5,2.5]$, and $U$ is the orthonormal matrix obtained by the singular value decomposition of a symmetric matrix $MM^T$, where the elements of the $p\times p$ matrix $M$ are uniformly distributed on $[-2,2]$.
In either cases of 5 or 20 clusters, we generated 10.000 vectors for each cluster, having dimensions $p=5,10,20$.

This procedure guarantees that these clusters are rather well-separated Gaussians, in particular for higher vector dimensions. 

\begin{table}
\begin{center}
{\tiny
\begin{tabular}{|l|l|l|l|l|l|l|}
\hline
n. of         & algorithm &  dimension $p$ & n. of data     & n. of               & n. of retained \\
true clusters & \         &  of data points & in each chunk & estimated   & points \\
\  & \ & \ & \ &  clusters &  (outliers)\\
\hline
5 & BFR & 5 & 25 & 6  & 0\\ \hline
5 & PA & 5 & 25 & 5  & 0\\ \hline
5 & BFR & 5 & 50 & 6  & 0\\ \hline
5 & PA & 5 & 50 & 5  & 0\\ \hline
5 & BFR & 10 & 25 & 5  & 0\\ \hline
5 & PA & 10 & 25 & 5  & 0\\ \hline
5 & BFR & 10 & 50 & 5  & 0\\ \hline
5 & PA & 10 & 50 & 5  & 0\\ \hline
5 & BFR & 20 & 25 & 5  & 0\\ \hline
5 & PA & 20 & 25 & 5 & 0\\ \hline
5 & BFR & 20 & 50 & 5 & 0\\ \hline
5 & PA & 20 & 50 & 5  & 0\\ \hline
20 & BFR & 10 & 25 & 12  & 0\\ \hline
20 & PA & 10 & 25 & 17  & 0\\ \hline
20 & BFR & 10 & 50 & 13  & 0\\ \hline
20 & PA & 10 & 50 & 22  & 1\\ \hline
20 & BFR & 20 & 25 & 11  & 0\\ \hline
20 & PA & 20 & 25 & 19 &  0\\ \hline
20 & BFR & 20 & 50 & 20  & 0\\ \hline
20 & PA & 20 & 50 & 20  & 0\\ \hline
\end{tabular}
\caption{Results of the application of our proposed algorithm (PA) and of the BFR algorithm to synthetic data. We call chunk the number of processed data out of which we apply secondary compression.}\label{results}
}
\end{center}
\end{table}

We applied both our procedure and the BFR algorithm to these synthetic data, to compare the performance of the two methods. In both cases, we computed the
secondary data compression once out of 25, or out of 50 data points. In the tests on data from 20 clusters we started from a lower number of initial clusters (equal to 10), in order to check the ability of our algorithm to detect the correct number of clusters. The results are reported in Table \ref{results}. 

We note that the number of clusters is sometimes underestimated by our method, in particular in the case of 20 clusters. In such cases, if the point clouds in different clusters are gathered in  rather close ellipsoids, then the correct detection of the clusters may be more difficult. Anyway in all cases the estimates provided by our algorithm are equal or better than those obtained with the BFR algorithm.

We also point out that in the case of 20 clusters with $p=10$, and secondary compression performed once out of 50 processed data, the overestimation of the number of clusters obtained with our algorithm is compensated by the presence of two small clusters, composed by a few hundreds of data points, which can then be revisited as groups of outliers. Anyway also in this case our results are better than those obtained with BFR.

The method seems to be sensitive to the frequency of the secondary compression only in presence of many clusters.

\begin{figure}[thb]
\begin{center}
\includegraphics[scale=0.4]{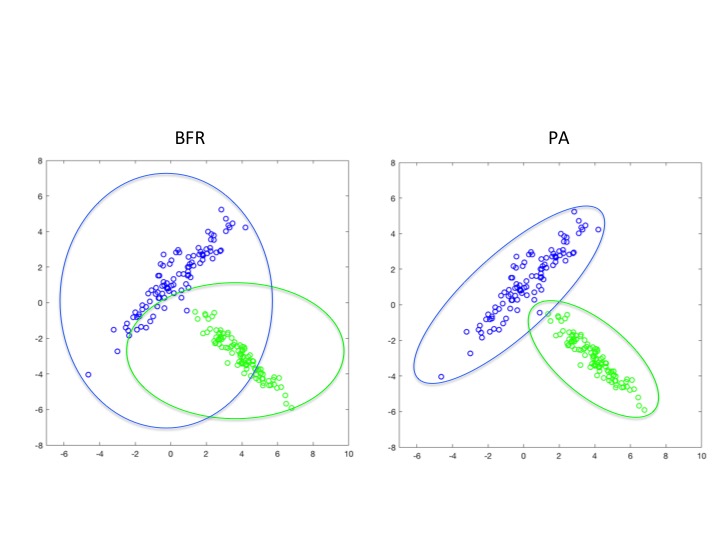}
\end{center}
\caption{Example of a typical situation in 2D: the BFR algorithm is approximating the shape of the clusters using ellipsoids parallel to the main axes. Our proposed algorithm (PA) is using ellipsoids with the correct orientation, thus the uncertainty region (overlapping of the ellipsoids) is reduced.}\label{fig:lowdim}
\end{figure}

Note that our method gives always a correct estimation of the number of clusters in all cases with 5 true clusters, while the BFR method overestimates the correct number in particular when the data dimension is small ($p=5$). This is reasonable since in lower dimensional spaces the  shape and orientation of the point clouds must be correctly estimated and taken into account to identify the clusters in a proper way (see Figure \ref{fig:lowdim}).

We tested also cases with bigger values of $p$, but in such cases both algorithms are able to detect the correct number of clusters, in an equivalent way, since a few clusters in high dimensional spaces are almost always well separated, because of "curse of dimensionality" reasons.

\subsection{Results on a real dataset}

We applied our algorithm to a real dataset to detect network intrusions.
Detecting intrusions is a typical data streaming problem, since it is essential to identify the event while it is happening.
In our experiments we used the KDD-CUP'99 (\texttt{http://kdd.ics.uci.edu/databases/kddcup99/kddcup99.html})   intrusion detection dataset which consists of two weeks of raw TCP dump data. This dataset is related to a local area network simulating a true Air Force environment with occasional attacks. Variables collected for each connection include the duration of the connection, the number of bytes transmitted from source to destination (and viceversa), the number of failed login attempts, etc. We applied our algorithm to the 34 variables that are declared to be continuous.

Some of these variables actually are almost constant, giving thus an estimated zero sample variance in many clusters. In such situation, if the BFR algorithm is applied, singular covariance matrices are estimated for some clusters. Consequently the Mahalanobis distance becomes unstable. Our optimal double shrinkage estimators are thus necessary to overcome this instability, and as a byproduct, they can take into account the deviation of the kurtosis from the Gaussian case.

The same dataset was analysed in \cite{O'Callaghan01streaming-dataalgorithms}, via the STREAM algorithm, but they used the Euclidean distance, which is a global distance that gives the same importance to all the variables.

We obtained stable results. We applied the secondary compression every 100 data, starting from 4 clusters composed by less than 20 points. We observed the presence of 6-8 big clusters starting from about 100000 processed data. We processed about 646000 data, ending with 5 big clusters, composed by the following number of points: 133028; 121661; 242206; 53235; 95977.
Note that we detected the final correct number of clusters, since in this dataset there are four possible types of attacks, plus no attacks. The four types of attacks are denial-of-service; unauthorized access from a remote machine (e.g. guessing password); unauthorized access to local superuser (root) privileges; surveillance and other probing (e.g., port scanning).

\begin{figure}[thb]
\begin{center}
\includegraphics[scale=0.25]{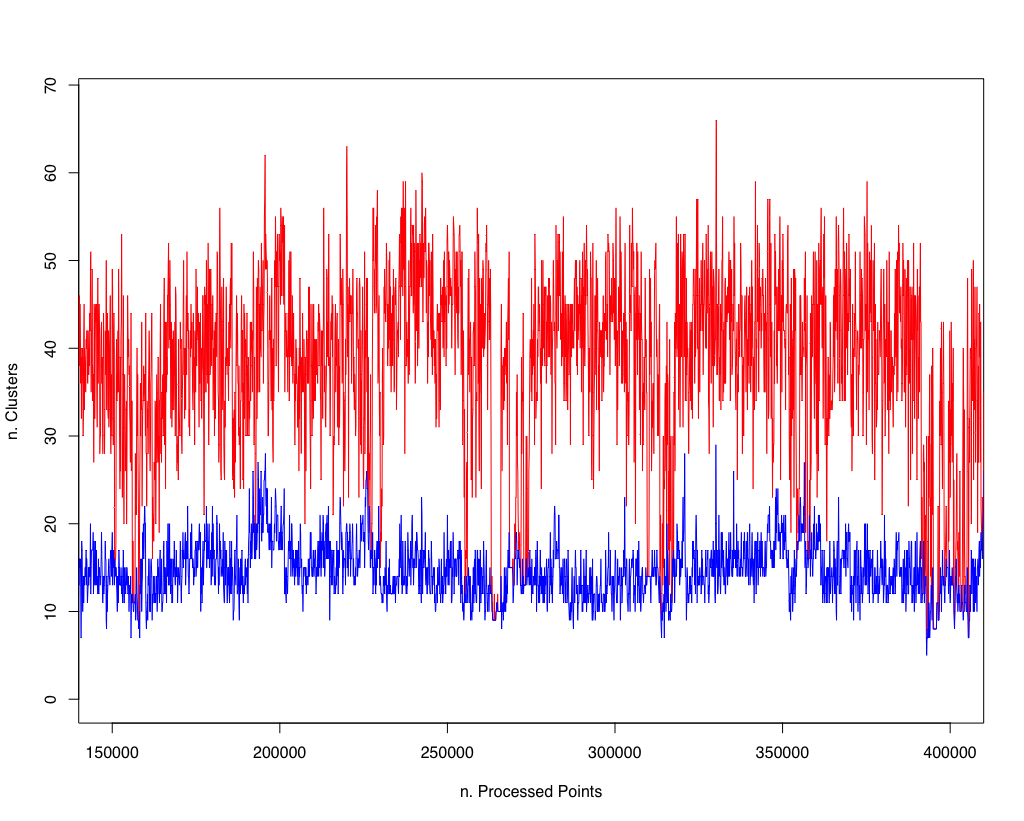}
\end{center}
\caption{Effect of the secondary compression. Red line: number of clusters obtained before secondary compression. Blue line: number of clusters obtained after secondary compression. Secondary compression is applied once out of 100 iterations.}\label{fig:experiment}
\end{figure}

In Figure \ref{fig:experiment} we show the effectiveness of secondary compression on the stabilization of the number of clusters. Actually when the number of identified clusters is bigger than 8 after secondary compression, the exceeding ones are formed just by a few points, and can then be reinterpreted as groups of outliers. For example when 300113 data have been processed, we find 15 clusters composed respectively by the following number of points:
95141; 22451; 50098; 79943; 30683; 11834; 7762; 1228; 712; 118; 100; 33; 4; 4; 2. Note that 7 out of 15 clusters are quite small, containing less than 1000 data points.

\section{Conclusion}

We have introduced a new algorithm to cluster data streams with correlated components. Our algorithm in some parts imitates the BFR algorithm, since, like BFR, it uses a local distance approach, based on the computation of the Mahalanobis distance. In order to compute such distance, positive definite estimators of the covariance matrices of the clusters are needed, also when the clusters contain just a few data points. We obtained such estimators by considering a Steinian double shrinkage method, which leads to covariance matrix estimators that are non-singular, well-conditioned, expressed in a recursive way and thus computable on data streams. Further such estimators provide positive definite estimates also when some components of the data points have a small variance, or the data distribution has a kurtosis different from the Gaussian case.

We applied both our proposed method and the BFR algorithm to synthetic gaussian data, and we compared their performance. From the numerical results we conclude that our method provides rather good clustering on synthetic data, and performs better than the BFR algorithm in particular in presence of few clusters in spaces of rather low dimension. This is reasonable since the BFR algorithm approximates the "clouds" of data with ellipsoids having axes parallel to the reference system, and this leads to a wrong classification when the clusters are elongated, not much separated, and with axes rotated with respect to the reference system. In such situations our algorithm is able to capture in a more proper way the geometry of the clusters, and thus improves the classification.

Anyway the secondary compression could be possibly improved by applying some incremental model-based technique (see \cite{Fraley2005}), but modified in such a way to avoid multiple scans of the sample.

We also applied our algorithm to a real dataset, obtaining good results in terms of correct identification of the number of clusters, and stability of our algorithm. 

The advantage of our algorithm with respect to other methods present in literature, like BFR or STREAM, is that it relaxes the  assumptions on the processed data streams, and can thus be effectively applied to a wider class of cases, on which it performs better. In the cases where the assumptions of the other methods are satisfied, our algorithm provides equivalent results. It can then be systematically substituted to other methods to analyze data streams, in all cases in which the data points are not too much high dimensional.

%\section{Conclusion}
%The extension of BRF to a framework which includes a ``full'' covariance matrix opens some issues that directly 
%join computational and statistical aspects.
%
%The Mahalanobis distance requires a matrix that must have positive determinant, and hence a shrinkage procedure has been chosen. The usual shrinkage methods cannot be directly updated

\section*{Declarations}

%\subsection*{Ethics approval and consent to participate}
%Not applicable
%
%\subsection*{Consent for publication}
%Not applicable
% 
%\subsection*{Availability of data and material}
%All data and material are available upon request.
%
%\subsection*{Competing interests}
%The authors declare that have no competing interests.

%\subsection*{Authors' contributions}
%The contribution of each author is equal.

%\subsection*{Acknowledgments }
%Not applicable

\appendix
\section{Proofs}

\noindent{\bf Proof of Lemma~\ref{lem:EQ_N}}

For $N=2$, as a consequence of the model:
\begin{multline*}%\label{eq:expQn}
E[\big( (\bx_2  - \bx_1 )^{\top} (\bx_2  - \bx_1 ) \big)^2] =
E[\big( (\by_2  - \by_1 )^{\top} (\by_2  - \by_1 ) \big)^2] \\
\begin{aligned}
& = E[(\by_2^{\top}\by_2)^2 ]  + 4 E[(\by_2^{\top}\by_1)^2 ] + E[(\by_1^{\top}\by_1)^2 ] + 
2 E[(\by_2^{\top}\by_2) (\by_1^{\top}\by_1)]
\\
& \qquad -2 E[(\by_2^{\top}\by_2)(\by_2^{\top}\by_1) ] -2 E[(\by_2^{\top}\by_1) (\by_1^{\top}\by_1)^2 ] 
\\
& = 2 E[(\by_1^{\top}\by_1)^2 ]  + 4 E[(\by_2^{\top}\by_1)^2 ] + 
2 E[(\by_1^{\top}\by_1) ]^2
\\
& \qquad -2 E[(\by_2^{\top}\by_2)\by_2^{\top} ] E[\by_1] -2 E[\by_2^{\top}] E[\by_1 (\by_1^{\top}\by_1)^2 ] 
\\
& = 2 E[(\by_1^{\top}\by_1)^2 ]  + 4 E[(\by_2^{\top}\by_1)^2 ] + 
2 E[(\by_1^{\top}\by_1) ]^2.
\end{aligned}
\end{multline*}
Since \( E[(\by_1^{\top}\by_1) ]= \tr \Sigma \), by \eqref{eq:Eyyyy} and \eqref{eq:Eyzyz}, we obtain
the first part of the thesis.

Let us add a point to a cluster of $N-1$ points. We obtain
\begin{multline*}%\label{eq:expQn}
E[Q_{N}] - E[Q_{N-1}] = E[Q_{N} - Q_{N-1}] 
\\
\begin{aligned}
&= E\big[ \big( (\bx_N  - \bxM^{(N)})^{\top} (\bx_N  - \bxM^{(N)}) \big)^2 \big] =
E\big[ \big( (\by_N  - \byM^{(N)})^{\top} (\by_N  - \byM^{(N)}) \big)^2 \big] \\
& = E[ (\by_N^{\top}\by_N - 2\by_N^{\top}\byM^{(N)} + \byM^{(N)\top}\byM^{(N)} )^2  ]
\\
& = \underbrace{E[ (\by_N^{\top}\by_N)^2]}_{A} + \underbrace{E[4 (\by_N^{\top}\byM^{(N)})^2]}_{B} +\underbrace{E[(\byM^{(N)\top}\byM^{(N)} )^2]}_{C} 
+ \underbrace{E[2 \by_N^{\top}\by_N \byM^{(N)\top}\byM^{(N)} ]}_{D} \\
& \qquad -\underbrace{E [ 4 \by_N^{\top}\by_N \by_N^{\top}\byM^{(N)}] }_{E}
-\underbrace{E [ 4 \byM^{(N)\top}\byM^{(N)} \by_N^{\top}\byM^{(N)}] }_{F}.
\end{aligned}
\end{multline*}
As above, the fact that $\by_n $ is independend from $\byM^{(n)}$, and
both have expectation null, imply 
\begin{align*}%\label{eq:eq2}
E & = 4 E [ \by_n^{\top}\by_n \by_n^{\top}\byM^{(n)}]
= 4 E [ \by_n^{\top}\by_n \by_n^{\top}] E [\byM^{(n)}] = 0 
\\
F & = E [ 4 \byM^{(n)\top}\byM^{(n)} \by_n^{\top}\byM^{(n)}]=
4 E [ \byM^{(n)\top}\byM^{(n)} \byM^{(n)\top} \by_n]=
4 E [ \byM^{(n)\top}\byM^{(n)} \byM^{(n)\top}] E[ \by_n]= 0
%\\ \label{eq:eq3}
\end{align*}
By \eqref{eq:Eyyyy},
\( A  = \kappa_{11} + 2 \tr (\Sigma^2) + (\tr \Sigma)^2 \).
By Lemma~\ref{lem:EyM2},
\( B  = \frac{4}{N-1} \tr (\Sigma^2) \).
By Lemma~\ref{lem:EMMMM},
\( C  =  
\frac{1 }{(N-1)^3}\kappa_{11} + \frac{2 }{(N-1)^2} \tr (\Sigma^2) + \frac{1 }{(N-1)^2}(\tr \Sigma)^2 \).
By Lemma~\ref{lem:EyyMM},
\( D  =  
\frac{2 }{N-1} (\tr \Sigma)^2 
\).
Then 
\begin{multline*}
E[Q_{N}] - E[Q_{N-1}]
= \Big(1 + \frac{1 }{(N-1)^3} \Big)\kappa_{11} + \Big(1 + \frac{2}{N-1} + \frac{1 }{(N-1)^2} \Big)
( 2\tr (\Sigma^2) + (\tr \Sigma)^2 )
\\
= \Big(1 + \frac{1 }{(N-1)^3} \Big)\kappa_{11} + \Big( 1 + \frac{1 }{N-1} \Big)^2
( 2\tr (\Sigma^2) + (\tr \Sigma)^2 ).
\end{multline*}

The case of merging two clusters is a simple consequence of \eqref{def:Q_N}, \eqref{def:S_N} and \eqref{def:T_N}.

\hfill{$\square$}

\end{document}